\pdfoutput=1
\documentclass[conference]{IEEEtran}
\IEEEoverridecommandlockouts
\usepackage{cite}
\usepackage{amsmath,amssymb,amsfonts}
\usepackage{algorithmic}
\usepackage{graphicx}
\usepackage{textcomp}
\usepackage{xcolor}
\usepackage{booktabs}
\usepackage{multirow} 
\usepackage{array}
\usepackage{subcaption} 
\usepackage{float}

\newcolumntype{L}[1]{>{\raggedright\arraybackslash}p{#1}}
\newcolumntype{C}[1]{>{\centering\arraybackslash}p{#1}}
\newcommand{\cm}{$\checkmark$}
\newcommand{\pcm}{($\checkmark$)}
\newcommand{\finding}[2]{\par\smallskip\noindent\textbf{Finding #1.}~\emph{#2}\par\smallskip}

\providecommand{\degree}{\ensuremath{^{\circ}}}   

\def\BibTeX{{\rm B\kern-.05em{\sc i\kern-.025em b}\kern-.08em
    T\kern-.1667em\lower.7ex\hbox{E}\kern-.125emX}}
\begin{document}

\title{Knit-Structure Effects on Electromechanical Metrics and Their Correlation with Joint-Angle Estimation Error in Knitted Strain Sensors}
\author{\IEEEauthorblockN{Annika Eloranta\textsuperscript{1},
Zhuchenyang Liu\textsuperscript{1},
Iiro Naulap\"{a}\"{a}\textsuperscript{1},
Iida Arvola\textsuperscript{1},
Yao Zhang\textsuperscript{1},\\
Anna-Mari Leppisaari\textsuperscript{1},
Lulu Xu\textsuperscript{2},
Yu Xiao\textsuperscript{1}}
\IEEEauthorblockA{\textsuperscript{1}\textit{Aalto University}, Espoo, Finland \qquad
\textsuperscript{2}\textit{Loughborough University}, Loughborough, UK}
}

\maketitle

\begin{abstract}
Knitted resistive strain sensors show strong promise for joint motion sensing in sports and rehabilitation, but the linkage between sensor design and in situ performance remains unclear. 
We investigate how knit structure and machine settings (e.g., stitch size) shape electromechanical properties and which metrics predict sensing performance during bending.
Sensors spanning seven common knit structures at two stitch sizes were fabricated, characterized under uniaxial cyclic tension, and evaluated on a joint‑emulating bending rig. Joint‑angle estimation was assessed with machine‑learning models, and correlations with electromechanical metrics were analyzed. Experimental results show that, among six common metrics, gauge factor and baseline resistance are largely set by knit structure, while working range, linear range, hysteresis, and cyclic stability vary only modestly across designs. Gauge factor correlates negatively and baseline resistance positively with joint‑angle estimation error—mainly in lower-sensitivity designs—whereas the other metrics have weak or no predictive value. These results support using uniaxial tensile tests to screen out weak designs, while underscoring the need for joint‑relevant evaluation and application‑specific metrics to identify top performers.
\end{abstract}


\section{Introduction}

Human motion sensing is a key technology for applications such as rehabilitation, sports training, ergonomics, and human--computer interaction~\cite{an2021mgait, sun2026textilesuit}. Body motion is well described by joint angles because movement primarily consists of rotations of body segments about anatomical joints, providing an interpretable and comparable representation across tasks and subjects~\cite{zhang2026lingomotion}. 
Knitted resistive strain sensors, which are stretchable and easy to integrate into textiles, can measure joint angles and their changes~\cite{kim2024measurement, schepers2026comparative}, making them 
well suited to interpretable body‑motion analysis.

Knit structure plays a central role in knitted strain sensor design: it sets the baseline deformation modes, governs how strain localizes and distributes, and shapes the evolution of conductive contacts during stretch. Consequently, the choice of structure, together with yarn selection and placement, directly influences electromechanical properties. %
Several studies have measured the electromechanical metrics listed in Table \ref{tab:metrics} for different knit structures (see Table \ref{tab:related}). 
However, heterogeneous test protocols and metric definitions, together with the limited set of structures evaluated, mean we still lack a systematic comparison across commonly used knit structures, yarn placements, and key machine settings (e.g., stitch size). Moreover, sensors are often selected based on metrics from uniaxial tensile tests, yet it remains unclear whether these metrics reliably predict body‑motion sensing performance, such as joint angle estimation error.

This work aims to answer two research questions (RQs): \textbf{RQ1:} How do knit structure and machine settings like stitch size influence the electromechanical characteristics of knitted resistive strain sensors? \textbf{RQ2:} Which electromechanical metrics from uniaxial tensile tests reliably predict joint angle estimation error for flexion–extension (bending) motions? We use bending motions as case studies, because they are common in functional tasks, provide a clear strain‑to‑angle mapping, and permit controlled, single‑degree‑of‑freedom evaluation of design effects.

To address RQ1, we fabricated sensor samples spanning seven knit structures commonly used in knitwear, each produced at two stitch sizes (14 designs total). We characterized their behavior via uniaxial cyclic tensile tests at three loading rates. We standardize sensor dimensions and yarn placement to ensure comparability. We then mounted the sensors on an elbow-emulating robot to assess how these electromechanical metrics translate to a joint-like setup. For RQ2, we trained machine‑learning models to estimate joint angles from sensor signals and empirically analyzed the correlation between electromechanical metrics and joint‑angle estimation error.

Below, we outline the key findings, which inform selection criteria for sensor designs suitable for motion analysis.

\begin{itemize}
\item Gauge factor (sensitivity) and baseline resistance are driven primarily by differences between knit structures; within‑design manufacturing variability is comparatively small. Specifically, these metrics vary 8–25$\times$ more across designs than across replicates. In contrast, working range, linear range, hysteresis, and cyclic stability vary at most 3$\times$ more across designs than within one, offering limited discrimination.
\item A one-step change in stitch size primarily tunes baseline resistance, with little effect on gauge factor. 
\item While tensile tests are useful for screening out weak designs, they are insufficient for selecting the best performers in joint-motion sensing. Of six metrics (Table \ref{tab:metrics}), only gauge factor and baseline resistance show any correlation with joint-angle estimation error, and only within the less-sensitive half of designs. The other metrics have weak or no predictive value.

\end{itemize}

For reproducibility, we will release the sensor designs, collected measurements, and the designs of the testing tools.

\section{Background and Related Works}

\subsection{Knitted Resistive Strain Sensors}

Knitted fabrics offer a flexible platform for strain sensing because their mechanical behaviour can be tailored through stitch structure, yarn properties and yarn placement \cite{atalay2014knitted, jansen2020performance}. Weft-knitted structures can be formed using knit, purl, tuck and float stitches, which can be combined into structures such as plain, rib, and interlock. Single- and double-bed knitting, together with techniques such as plating, inlay and stitch transfer, provide further possibilities for controlling fabric geometry and conductive-yarn placement. The resulting properties depend on the interaction between knitting gauge, structure, yarn material, yarn tension, stitch dimensions and other knitting parameters \cite{atalay2014knitted, jansen2020performance}. These parameters influence both fabric deformation and the electrical response of conductive yarns incorporated into the structure.

Knitted strain sensors typically pair conductive with elastic or nonconductive yarns, integrating the conductive yarn into the fabric or controlling its path via plating or inlay \cite{atalay2014knitted, oh2024evaluation}. Deformation alters loop geometry, conductive‑path length, and inter‑segment contacts, thereby changing resistance. The response thus reflects coupled knit mechanics and the evolving conductive network \cite{atalay2014knitted, jansen2020performance}, with the balance set by knit structure, the conductive‑yarn system, and the magnitude and mode of deformation.

Design spaces for knitted sensors therefore span knit structure (e.g., plain, rib, interlock), conductive-yarn placement (e.g., plated vs inlaid), yarn tension and thickness, the number and arrangement of conductive courses/wales, and overall knitting density \cite{atalay2014knitted, oh2024evaluation}. By tuning these parameters, designers can target sensitivity, working range, and signal linearity. Meanwhile, knitted sensors often exhibit nonlinearity, hysteresis, viscoelastic/time-dependent effects (e.g., creep, relaxation), and recovery dynamics, making preconditioning, repeatability, and stabilization critical for practical performance \cite{jansen2020performance}.

\subsection{Electromechanical Metrics}
\label{sec:metrics}

\begin{table}[t]
\caption{Metric definitions. WR, LR, GF and H are computed on the cycle-averaged loading curve of $\Delta R/R_0$ against strain. A window is one thirtieth of the stroke.}
\label{tab:metrics}
\centering
\footnotesize
\begin{tabular}{p{2.0cm}p{5.9cm}}
\hline
Metric & Definition \\
\hline
Baseline resistance $R_0$ & The resistance of the unstrained sensor, which is the reference for the relative change $\Delta R/R_0$~\cite{liu2023review}.
  \\
Working range (WR) & The longest strain interval in which $\Delta R/R_0$
  increases by more than $0.002$ over every window~\cite{schepers2026comparative}. \\
Linear range (LR) & The longest interval within the WR in which a linear fit
  reaches $R^2 \geq 0.96$~\cite{schepers2026comparative}. \\
Gauge factor (GF) & The slope of this linear fit, expressed as $\Delta R/R_0$
  per unit strain~\cite{schepers2026comparative}. \\
Hysteresis (H) & The largest difference in strain between the loading and
  unloading curves at equal $\Delta R/R_0$ within the WR~\cite{schepers2026comparative}. \\
Cyclic Stability (CA) & The standard deviation of $R$ across cycles at each
  point of the cycle, averaged over the cycle~\cite{liu2023review}.  \\
\hline
\end{tabular}
\end{table}

Knitted strain sensors are commonly characterized by uniaxial tensile testing,
in which a sample is stretched and released between two clamps of a texture
analyser or a linear stage while its resistance is recorded through a
voltage divider~\cite{warncke2023advancing}. %
Table~\ref{tab:metrics} summarizes the  metrics commonly used in tensile testing and the calculation rules adopted in this study.

\textbf{Working Range (WR)} is the range of strain over which one resistance value
corresponds to one strain value~\cite{jansen2020performance}.  \textbf{Linear Range (LR)} is the part
of the working range in which the resistance change is linear in
strain~\cite{bozali2022development}, while \textbf{Gauge Factor (GF)} is the slope of this linear
relation, which quantifies the sensitivity of the sensor~\cite{liu2023review}.
\textbf{Hysteresis (H)} describes the difference between the stretch and release directions of a cycle; a low hysteresis is required so that a joint angle can be determined independently of the previous position of the
joint~\cite{bozali2022development}. %
\textbf{Cyclic Stability (CA)} is assessed by repetitive stretch–release at constant speed and quantified as the standard deviation around the cycle‑averaged response~\cite{liu2023review}.

\subsection{Related Works}
\label{sec:relatedwork}

\begin{table*}[t]
\caption{Comparison with prior experimental studies of knitted resistive strain sensors. \cm{} reported, \pcm{} partially or qualitatively, -- not reported. Samples: samples tested per design; Multiple rates: more than one loading rate tested; Bend/body testing: bending rig or on-body test in addition to tensile testing; Angle estimation: joint angle estimated from the sensor signal against a reference angle; Correlation analysis: tensile-test metrics correlated with joint angle estimation error.}
\label{tab:related}
\centering\footnotesize\setlength{\tabcolsep}{3pt}
\begin{tabular}{l L{3.0cm} C{1.1cm} C{1.0cm} cccccc C{1.3cm} C{1.3cm} C{1.5cm}}
\toprule
 & & & & \multicolumn{6}{c}{Metrics reported} & & & \\
\cmidrule(lr){5-10}
\multirow{-2}{*}{Study} & \multirow{-2}{=}{Knit structures} & \multirow{-2}{=}{Samples} & \multirow{-2}{=}{Multiple rates} & $R_0$ & GF & WR & LR & H & CS & \multirow{-2}{=}{Bend/body testing} & \multirow{-2}{=}{Angle estimation} & \multirow{-2}{=}{Correlation analysis} \\
\midrule
Atalay \& Kennon~\cite{atalay2014knitted}   & interlock (1)                        & 1     & --   & \cm & \cm & \cm & \cm & \cm & --  & --   & -- & -- \\  %
Atalay et al.~\cite{atalay2017comparative}  & plain, interlock (2)                 & 5     & --   & --  & \cm & \pcm & \cm & --  & \pcm & --   & -- & -- \\
Isaia et al.~\cite{isaia2020study}          & jersey with tuck and miss stitches (1)  & 2--3  & --   & --  & \pcm & --  & \cm & --  & \cm & \cm  & -- & -- \\  %
Bozali et al.~\cite{bozali2022development}  & 1$\times$1 rib (1)                   & 1--3  & \pcm & \cm & \cm & \cm & \pcm & \cm & \pcm & --  & -- & -- \\  %
Warncke et al.~\cite{warncke2023advancing}  & plain, rib, interlock, purl (4)      & --  & --   & \cm & \cm & \pcm & \cm & \cm & \pcm & --  & -- & -- \\  %
Fischer et al.~\cite{fischer2024novel}      & plain (1)                            & 5     & --   & \cm & \cm & \cm & \pcm & \pcm & \pcm & --  & -- & -- \\  %
Kim et al.~\cite{kim2024assessing}          & plain, 2 plating sides (1)           & --  & --   & \cm & \cm & \cm & \cm & \pcm & \cm & \cm  & -- & -- \\  %
Oh \& Kim~\cite{oh2024evaluation}           & plain, purl, rib (3)                 & 1     & \pcm & --  & \cm & --  & --  & --  & \cm & \cm  & \cm & -- \\  %
Schepers et al.~\cite{ schepers2026comparative} & plain (1), 10 patterns & 1 & \cm & \cm & \cm & \cm & \cm & \cm & \cm & \cm & \cm & -- \\  %
Abbas et al.~\cite{abbas2026tuning}         & jersey, cross-tuck, cross-miss (3)   & --  & \pcm & \cm & \cm & \pcm & --  & \cm & --  & \cm  & -- & -- \\  %
\midrule
\textbf{This work} & \textbf{7 structures $\times$ 2 stitch sizes} & \textbf{4} & \cm & \cm & \cm & \cm & \cm & \cm & \cm & \cm & \cm & \cm \\
\bottomrule
\end{tabular}
\end{table*}

Previous studies (Table \ref{tab:related}) often focus on the electromechanical properties of a single knit structure and analyze trade-offs among those properties. For example, in plated plain-knit sensors, authors have reported GF above 30 with LR of 10–30\,\% of strain~\cite{li2021sensing, oh2024evaluation} versus GF below 2 with LR of 40 to 50\,\% of strain\cite{bozali2022development, fischer2024novel}. Recently, Oh and Kim \cite{oh2024evaluation} compared plain, purl, and rib structures and investigated plating, yarn thickness, and needle position, demonstrating that both stitch structure and conductive-yarn placement substantially affect sensor sensitivity and electrical characteristics. Schepers et al.~\cite{schepers2026comparative} compared ten plain-knit patterns in terms of working range, linear range, gauge factor, and hysteresis, using speed, hold, and cyclic stability tests. This paper extends that line of work by examining 7 knit structures at different stitch sizes and by analyzing between‑sample variation, strain‑rate effects, and correlations with achievable joint‑angle sensing performance.

Among the few studies \cite{oh2024evaluation, schepers2026comparative} that conducted bending tests, Oh and Kim \cite{oh2024evaluation} considered only $60\degree$, $90\degree$, and $120\degree$, while Schepers et al. \cite{schepers2026comparative} posed joint‑angle estimation as a classification problem, limiting resolution to $10\degree$. In contrast, we analyze continuous bending and perform regression‑based joint‑angle estimation with higher resolution and accuracy.

\begin{table*}[t]
\caption{Description of the 14 designs. Needle counts are reported for the front and back beds. Course and wale counts refer to the sensing region; course counts are given for the smaller and larger stitch sizes, respectively.}
\label{tab:structures}
\centering\footnotesize\setlength{\tabcolsep}{3pt}
\begin{tabular}{lllp{3.4cm}p{2.1cm}llr}
\toprule
Knit structure & Abbreviated name & Beds, needles & Repeat in the sensing region & Conductive yarn & Stitch sizes & Courses & Wales \\
\midrule
Plain jersey & Plain jersey (PJ) & 1, 24 & all needles knit & plated & 10.5, 11 & 170, 150 & 24 \\
1x1 rib & 1x1 rib (Rib) & 2, 16+16 & all needles knit & plated & 10, 10.5 & 140, 130 & 32 \\
1x1 full cardigan & Full cardigan (FC) & 2, 12+12 & knit and tuck stitches in alternating rows  on both beds& plated & 10, 10.5 & 180, 170 & 24 \\
1x1 half cardigan & Half cardigan (HC) & 2, 12+12 & knit stitches in the front bed, tuck stitches in every second row on the back bed& plated & 10, 10.5 & 180, 170 & 24 \\
2x1 rib & 2x1 rib (2x1R) & 2, 18+18 & all needles knit & plated & 10, 10.5 & 100, 100 & 36 \\
Jersey with conductive inlay & Cond.\ inlay (CI) & 1, 33 & 2 courses jersey, 2 courses with knit on 2 needles and floating over 3 needles& on the float rows plated& 10.5, 11 & 200, 198 & 33 \\
Tuck-enhanced rib (1x1) & Tuck rib (TR) & 2, 17+17 / 15+15 & 4 courses knit, 1 tuck course on every other needle& plated & 10, 10.5 & 140, 133 & 34 \\
\bottomrule
\end{tabular}
\end{table*}

\section{Sensor Fabrication}
\label{sec:materials}
\begin{figure}[t]
\centering
\includegraphics[width=\columnwidth]{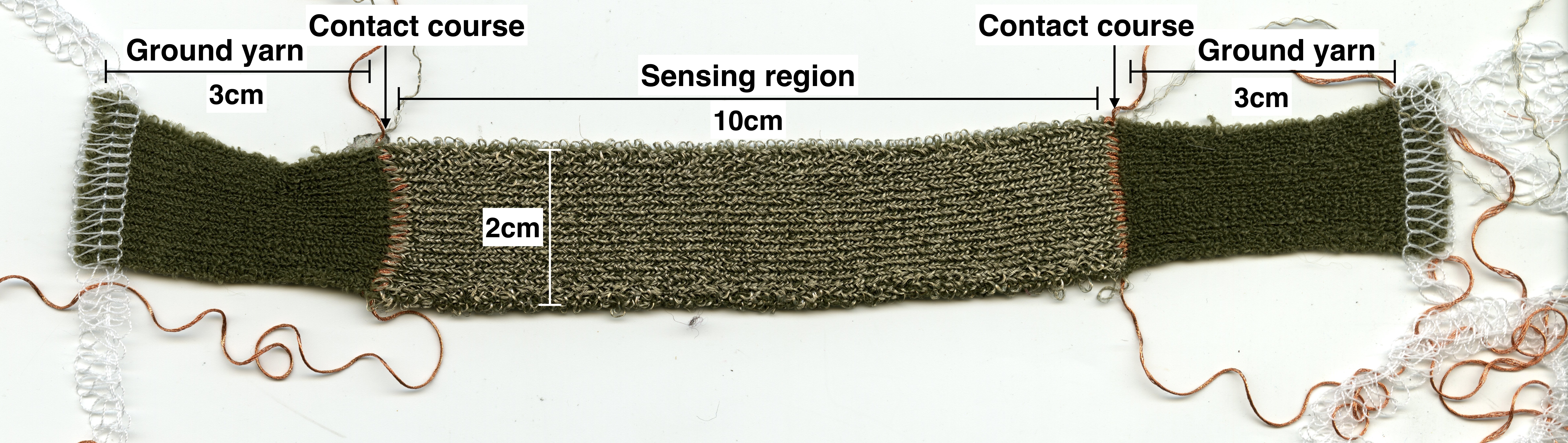}
\caption{Illustration of yarn placement.}
\label{fig:samples}
\end{figure}

Seven knit structures (Table~\ref{tab:structures}) were produced, each at two adjacent stitch sizes, yielding fourteen designs. Here stitch size refers to the machine setting that controls loop length. For each design, four samples were knitted. Three samples underwent uniaxial tensile testing; one of these was subsequently measured on the bending testing rig together with a fourth, previously untested sample (Section~\ref{sec:protocol_measurement}).

As illustrated in Figure \ref{fig:samples}, all samples share the same layout and yarn placement. Specifically, a 3\,cm margin of ground yarn, one contact course, a sensing region of 2\,cm by 10\,cm after steaming, a
second contact course and a 3\,cm margin. The ground yarn is Baktron 1300
(Nm 53.7/1, 90\,\% polyester, 10\,\% Lycra);  the conductive yarn is Shieldex 117/17 dtex Z100 HCB, which is present only in the sensing region and in the two contact courses, where it is knitted together with a copper yarn that forms the electrodes.

All samples were knitted on a Dubied 12\,gg hand-operated flat knitting machine. The number of beds is determined by the chosen knit structure. Among the seven structures, two are single-bed: plain jersey and jersey with a conductive inlay. In the inlay variant, two courses of jersey alternate with two courses in which only two needles knit and the conductive yarn floats across three needles, so the conductive yarn lies on the back of the fabric. The remaining five are double-bed rib structures: 1×1 rib, 2×1 rib, 1×1 full cardigan, 1×1 half cardigan (the latter two are 1×1 ribs incorporating tuck stitches on both beds or on the back bed only), and a tuck‑enhanced rib (a 1×1 rib with one tuck course on every other needle after every four knit courses). In all structures except
the Jersey with conductive inlay, the conductive yarn is plated onto the ground yarn in the sensing region, so that it lies on one face of the fabric. 

The number of needles engaged corresponds to the wale count. In this study, the sensor sample dimensions were fixed, thus the number of needles differs across structures. To achieve comparable fabric tightness and stable knitting, we adjusted stitch size by structure: plain jersey and the inlay were knitted at sizes 10.5 and 11, while the rib-based structures were knitted at 10 and 10.5.
The contact courses were knitted at a larger
stitch size (11 or 11.5) than the sensing region.

\section{Experimental Setup}

For all designs, we performed uniaxial tensile and bending tests. Section \ref{sec:tools} describes the testing tools, while Sections \ref{sec:protocol_measurement} and \ref{sec:analysis_protocol} present the test and analysis protocols.

\subsection{Testing Tools}
\label{sec:tools}

\begin{figure}[t]
\centering
\scalebox{0.7}{\includegraphics[width=\columnwidth]{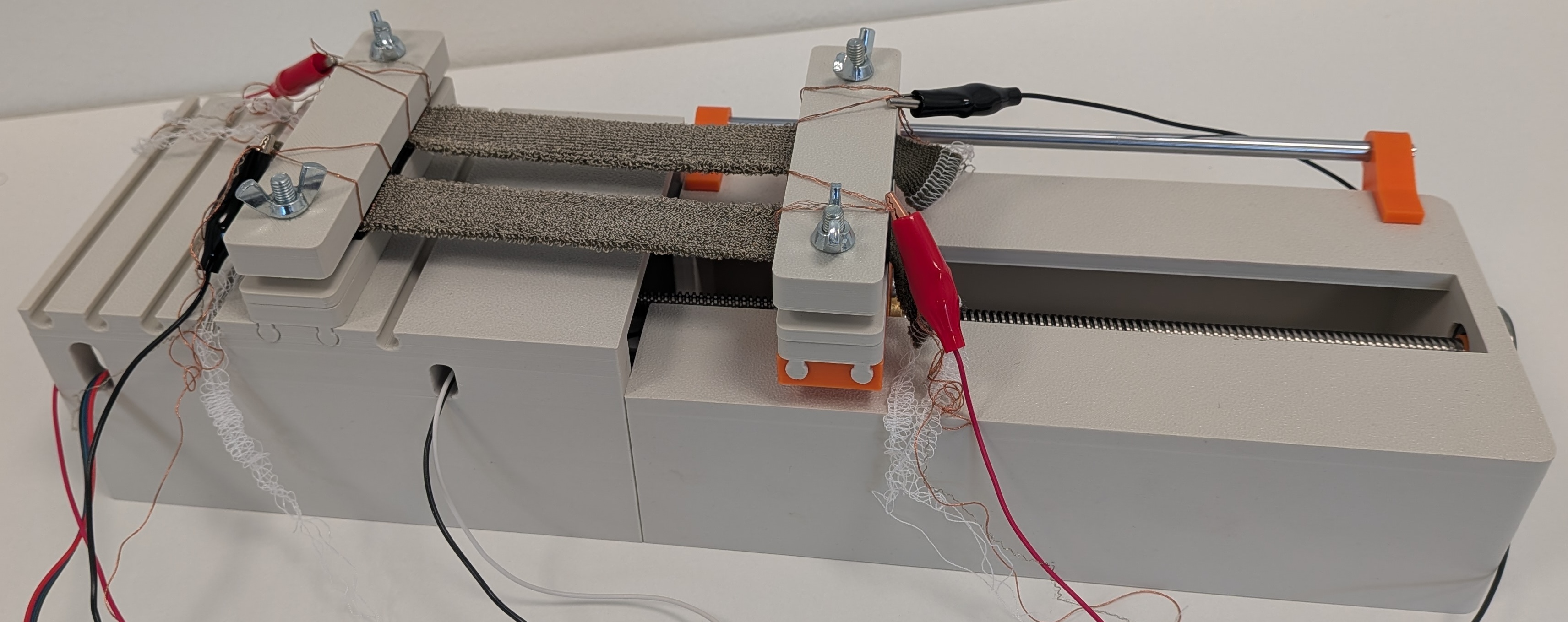}}
\caption{Tensile test setup}
\label{fig:tensile-test-tool}
\end{figure}

\textbf{Tensile Test:} We performed tensile tests on a custom-built platform (Figure \ref{fig:tensile-test-tool}) comprising both 3D‑printed and mechanical components. Linear motion was provided by a stepper‑motor–driven lead screw, enabling precise control of stretch distance and speed. Custom clamps prevented slippage; samples were gripped at both ends, with one clamp attached to the actuator.
The electronics comprised of an Arduino Nano, a DRV8825 driver, a NEMA 17 stepper motor, and two voltage‑divider channels to measure up to two samples simultaneously. Each sample was connected via crocodile clips to a known reference resistor; the Arduino's 10-bit ADC read the divider midpoint to compute resistance.

\begin{figure}[t]
\centering
\includegraphics[height=0.15\textheight]{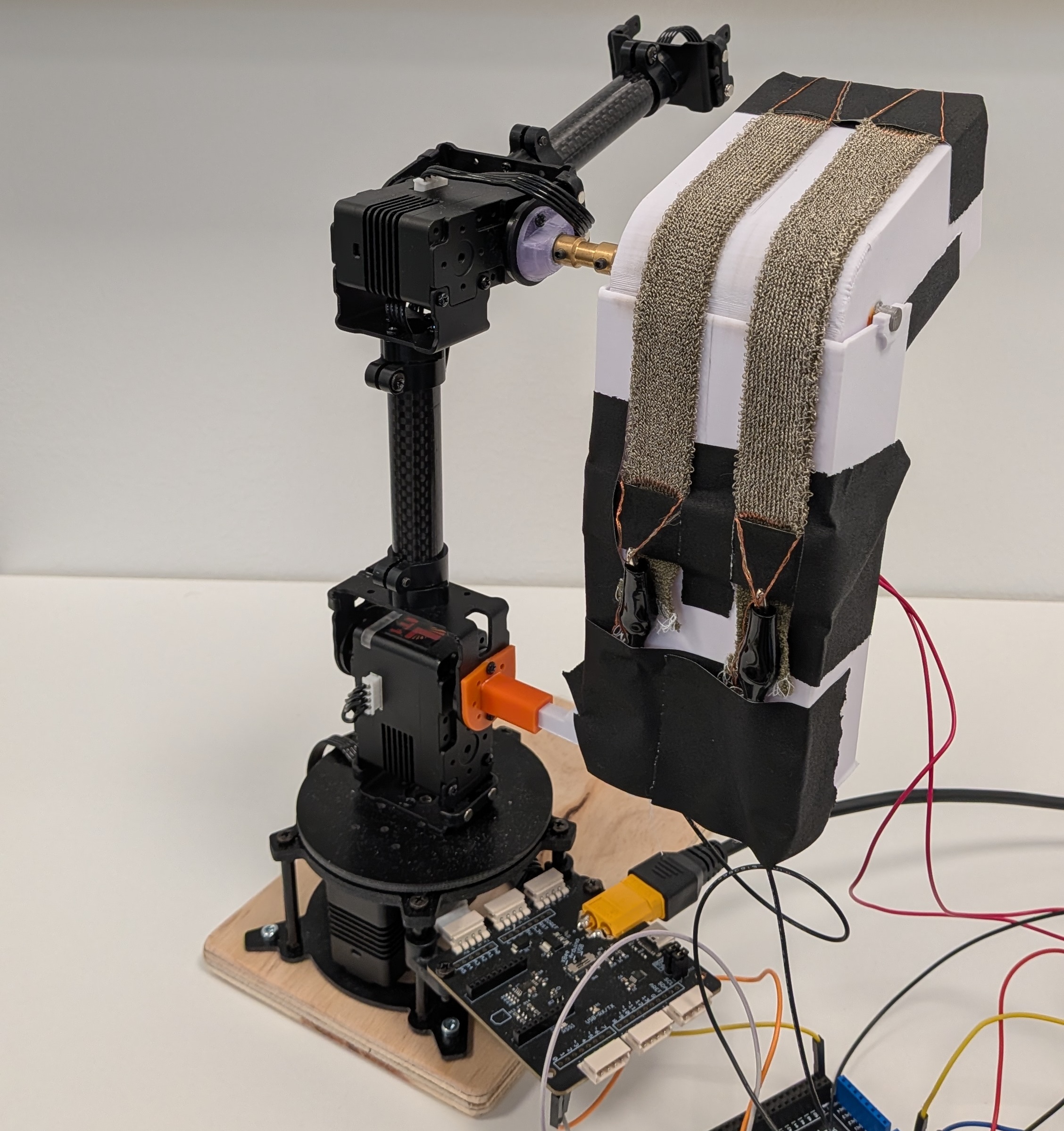}
\caption{Bending test setup}
\label{fig:bending-test-tool}
\end{figure}

\textbf{Bending Test:} We 3D‑printed an upper arm, elbow, and forearm, and actuated them using a Lynxmotion SES‑V2 robotic arm with two attachment points (Figure \ref{fig:bending-test-tool}). The 3D‑printed assembly was synchronized with the robot arm's movement through a connecting axle. Communication with the robot arm was performed using an Arduino Mega 2560 Rev3, enabling precise control of joint angle, angular velocity, and cycle count. Furthermore, a voltage divider circuit was utilized to sample the resistance of the sensor. Each sensor sample was mounted with the centre of the sensing region aligned with the centre of rotation. The sensor ends were secured with Velcro and gaffer tape, while the sides remained free to allow the knit to deform naturally. 

In both tensile and bending tests, resistance and extension (or joint angle) were sampled at 100 Hz and streamed to the laptop as CSV over serial.

\begin{figure*}[t]
\centering
\includegraphics[width=\textwidth]{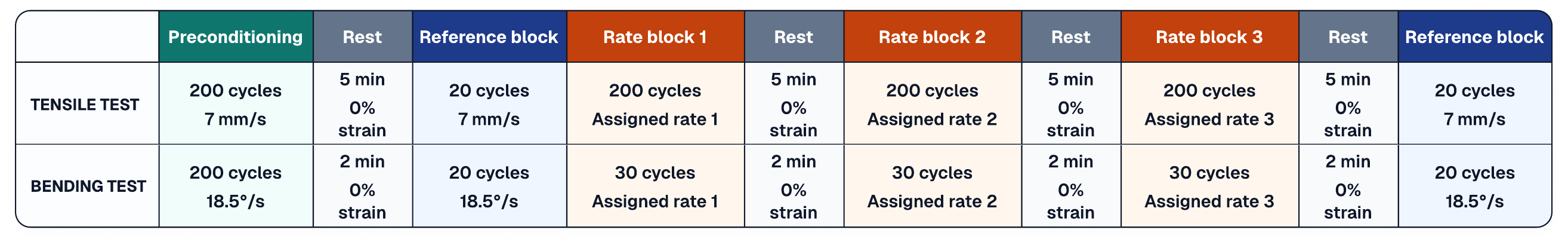}
\caption{Schematic overview of the tensile and bending test  protocol.}
\label{fig:protocol}
\end{figure*}

\subsection{Testing Protocol}
\label{sec:protocol_measurement}

\textbf{Tensile Test:} The knitted strain sensors were characterized using uniaxial cyclic tensile testing over $0$--$60\%$ engineering strain. Three loading rates were tested: $3$, $7$, and $15  \mathrm{mm/s}$, with identical stretch and release speeds within each condition. To mitigate order effects, the three samples of each design were assigned different rate sequences using a Williams trio counterbalance: $3$--$7$--$15$, $7$--$15$--$3$, and $15$--$3$--$7$. The medium rate block appeared once in each session position (first, second, third), yielding design means balanced with respect to elapsed time within the session. 

The measurement sequence is illustrated in Figure \ref{fig:protocol}. Each specimen was preconditioned for 200 cycles at $7\mathrm{mm/s}$ to $60\%$ strain, followed by a 5‑min unloaded rest. Each measurement session comprised five blocks: (1) a 20‑cycle reference at $7\mathrm{mm/s}$; (2) three rate blocks (one per loading rate), 200 cycles each, with 5‑min unloaded rests between blocks; and (3) a second 20‑cycle reference at $7\mathrm{mm/s}$.

Before each measurement, the gauge length of the sample $L_0$ was recorded at zero strain. Engineering strain was computed as $\varepsilon = (L - L_0)/L_0$; thus, $60\%$ strain corresponds to elongation to $160\%$ of the original length. Throughout, specimens were stretched to $160\%$ of $L_0$ and then released back to zero strain. After each session, the unloaded specimen length was compared with its initial value to assess permanent elongation, and any visible deformation or damage was documented.

\textbf{Bending Test: } 
The measurement protocol mirrored the tensile tests, as can be seen in Figure \ref{fig:protocol}. Specimens not previously tensile-tested were preconditioned for 200 cycles at the medium rate. Each session comprised a 20‑cycle reference at the medium rate, followed by three rate blocks of 30 cycles each, with 2‑min unloaded rests between blocks, and concluded with a 20‑cycle reference at the medium rate. Unlike the tensile tests, all previously untested samples followed a fixed rate order with the medium block in the third position, ensuring that all fourteen designs are compared at the same medium rate and at the same point in the session.

Let $\theta$ denote the joint angle and $L_0$ the length of the mounted sample at $\theta = 0\degree$. The calibration of the robotic arm (Section~\ref{sec:protocol_measurement}) established that a flexion from $0\degree$ to $150\degree$ lengthens the sensor sample by $\Delta L = 61$\,mm. Assuming elongation varies proportionally with angle, the strain at angle $\theta$ is
\begin{equation}
\varepsilon(\theta) = \frac{\Delta L}{L_0}\cdot\frac{\theta}{150\degree},
\end{equation}
which is the strain definition of Section~\ref{sec:protocol_measurement} applied to the calibrated elongation. Because calibration was performed only at the endpoints, the linearity between them is assumed; all strain-derived quantities for the bending test rig inherit this assumption. Under the same calibration, angular rates of $8.3$, $18.5$ and $39.5\degree/\mathrm{s}$, correspond to elongation rates
of $3.4$, $7.5$ and $16$\,mm/s, respectively, aligning the slow, medium and fast blocks across the two rigs.

Room temperature and relative humidity were recorded at the start and end of each measurement session in both tensile and bending tests.

\subsection{Analysis Protocols}
\label{sec:analysis_protocol}

For each session, metrics were first computed per block (five blocks per session). To avoid rate-dependent confounds and ensure consistency across both rigs and reference measurements, sample-level values were taken exclusively from the medium-rate block rather than averaged across rates. Design-level metrics are reported as the arithmetic mean and variance across independent samples of the same design.

\textbf{Tensile Test: }
For each block, $R_0$ was defined as the mean resistance over the second half of the preceding rest period, since knitted sensors relax during rest and exhibit drift under repeated loading~\cite{liu2023review}.
Resistance within the block was then expressed as $\Delta R/R_0$. Of
the 200 cycles of a rate block, the first 50 were discarded to allow for settling after mounting and rate changes;
all 20 cycles of a reference block were retained. The remaining
cycles were resampled onto a
strain grid of one six-hundredth of the stroke and averaged separately for loading and unloading.
 
From the averaged loading and unloading curves, WR, LR, GF and H were computed according
to the definitions in Table~\ref{tab:metrics}. The sliding window of the
WR rule is one thirtieth of the stroke and its threshold is
$0.002$ per window, equivalent to the 20-sample window and the
$10^{-4}$-per-sample threshold of Schepers et al.~\cite{schepers2026comparative} at
their sampling rate and speed; the increase is measured between the mean
values of consecutive windows. CA, in contrast, was computed from the individual retained
cycles rather than from the averaged curve.
Because the resistance also drifts monotonically over the cycles of a
block, a linear trend across cycles was removed at each point of the cycle
before the standard deviation was taken.

\textbf{Bending Test: }
Measurements from the bending tests were analyzed as in the tensile tests, substituting joint angle for strain: data were resampled onto a $0.25\degree$ grid, and a $5\degree$  sliding window was applied over the $150\degree$ stroke.

\section{Electromechanical properties}
\label{sec:results_em}

\begin{figure*}[t]
\centering
\includegraphics[width=\textwidth]{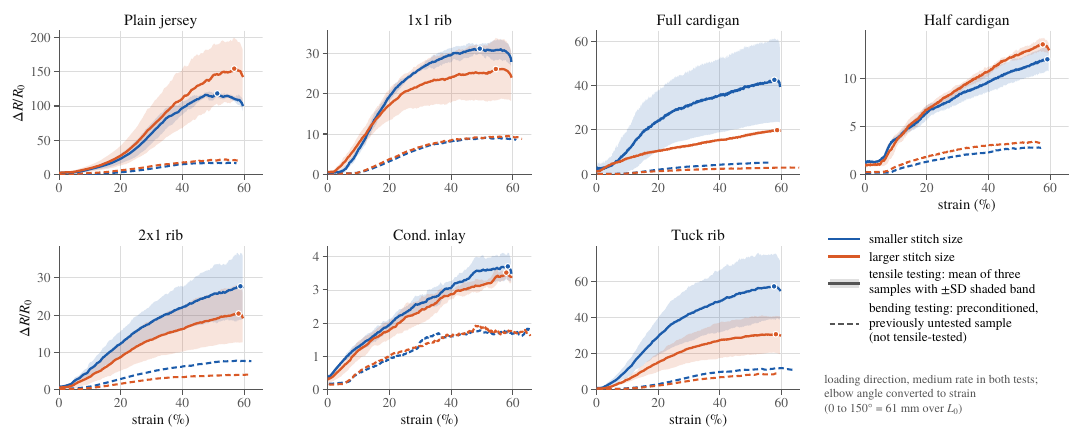}
\caption{Comparison of resistance–strain curves from uniaxial tensile and bending tests. The y-axis shows cycle-averaged resistance at each strain level (x-axis). Stitch sizes are distinguished by color. Solid lines are the mean tensile-test curves across three samples at the medium rate; shaded bands indicate $\pm$SD, and the dot marks the peak of the mean curve. Dashed lines are measurements from the previously unused sample in the bending test at 18.5\degree/s, with joint angle converted to strain via the robot calibration.}
\label{fig:curves}
\end{figure*}

Table~\ref{tab:designs} and Figure~\ref{fig:curves} compare the electromechanical metrics
across the fourteen designs. The remainder of this section presents the electromechanical properties measured in the tensile and bending tests and assesses their consistency.

\begin{table*}[t]
\caption{Electromechanical properties of the fourteen designs.  Tensile tests: medium rate (7\,mm/s), mean$\pm$SD over three samples per design.  Bending tests: medium rate (18.5\degree/s), angle converted to strain, mean$\pm$SD over two samples per design.  All measurements were performed at room temperature of 21.3--23.7\,$^\circ$C and relative humidity of 50--63\%.  GF is $\Delta R/R_0$ per unit strain; WR, LR and H in \% strain; Stab.\ is the detrended cycle-to-cycle SD in $\Omega$. Arrows indicate the conventionally preferred direction; the best value in each column is bolded ($R_0$ has no preferred direction).  Last row: between-design SD divided by the median within-design SD on the tensile testing rig.}
\label{tab:designs}
\centering\footnotesize\setlength{\tabcolsep}{2pt}
\begin{tabular}{ll|rrrrrr|rrrrrr}
\toprule
 & Stitch & \multicolumn{6}{c|}{Tensile tests ($n=3$)} & \multicolumn{6}{c}{Bending tests ($n=2$)} \\
Structure & size & $R_0$ ($\Omega$) & GF$\uparrow$ & WR$\uparrow$ & LR$\uparrow$ & H$\downarrow$ & Stab.$\downarrow$ & $R_0$ ($\Omega$) & GF$\uparrow$ & WR$\uparrow$ & LR$\uparrow$ & H$\downarrow$ & Stab.$\downarrow$ \\
\midrule
Plain jersey & 10.5 & 1.23$\pm$0.05 & 302$\pm$14 & 47.6$\pm$3.0 & 44.1$\pm$6.1 & \textbf{4.5$\pm$0.5} & 1.96$\pm$0.20 & 2.61$\pm$0.31 & \textbf{46$\pm$2} & 45.6$\pm$0.7 & 45.6$\pm$0.7 & \textbf{7.4$\pm$2.5} & 0.36$\pm$0.02 \\
 & 11 & 0.71$\pm$0.33 & \textbf{329$\pm$107} & 54.9$\pm$2.6 & 54.9$\pm$2.6 & 10.6$\pm$1.7 & 1.40$\pm$0.30 & 1.92$\pm$0.05 & 42$\pm$9 & 50.0$\pm$1.3 & 49.9$\pm$1.1 & 11.9$\pm$0.1 & 0.31$\pm$0.08 \\
\addlinespace[1pt]
1x1 rib & 10 & 1.01$\pm$0.13 & 91$\pm$3 & 39.9$\pm$6.1 & 37.5$\pm$3.9 & 10.5$\pm$2.3 & 0.92$\pm$0.01 & 2.18$\pm$0.30 & 25$\pm$1 & 47.3$\pm$6.1 & 42.6$\pm$0.6 & 11.0$\pm$2.5 & 0.13$\pm$0.04 \\
 & 10.5 & 1.54$\pm$0.43 & 77$\pm$13 & 39.5$\pm$10.7 & 32.6$\pm$5.8 & 12.4$\pm$4.5 & 0.96$\pm$0.03 & 3.27$\pm$1.25 & 23$\pm$1 & 39.4$\pm$6.7 & 36.1$\pm$11.3 & 13.2$\pm$0.9 & 0.19$\pm$0.08 \\
\addlinespace[1pt]
Full cardigan & 10 & 0.32$\pm$0.22 & 67$\pm$47 & 56.7$\pm$1.3 & 48.0$\pm$6.7 & 6.8$\pm$4.0 & 0.81$\pm$0.36 & 1.64$\pm$0.12 & 8$\pm$4 & 55.4$\pm$1.1 & 48.1$\pm$9.3 & 10.0$\pm$4.7 & 0.07$\pm$0.00 \\
 & 10.5 & 0.47$\pm$0.06 & 28$\pm$1 & \textbf{58.7$\pm$0.1} & \textbf{58.7$\pm$0.1} & 8.6$\pm$2.2 & 0.73$\pm$0.27 & 1.30$\pm$0.00 & 7$\pm$0 & \textbf{60.8$\pm$5.3} & \textbf{55.2$\pm$13.2} & 14.9$\pm$3.9 & \textbf{0.06$\pm$0.01} \\
\addlinespace[1pt]
Half cardigan & 10 & 1.08$\pm$0.09 & 18$\pm$2 & 57.6$\pm$0.3 & 56.2$\pm$2.6 & 13.2$\pm$1.2 & 0.70$\pm$0.33 & 2.80$\pm$0.27 & 6$\pm$0 & 48.6$\pm$0.4 & 42.7$\pm$8.8 & 18.3$\pm$4.4 & 0.08$\pm$0.01 \\
 & 10.5 & 1.15$\pm$0.16 & 23$\pm$3 & 57.1$\pm$0.3 & 56.0$\pm$2.1 & 13.1$\pm$1.0 & \textbf{0.68$\pm$0.33} & 2.61$\pm$0.00 & 8$\pm$0 & 48.5$\pm$2.4 & 41.4$\pm$6.3 & 18.0$\pm$0.1 & 0.08$\pm$0.00 \\
\addlinespace[1pt]
2x1 rib & 10 & 1.37$\pm$0.43 & 52$\pm$14 & 53.3$\pm$4.3 & 52.8$\pm$5.0 & 10.7$\pm$1.9 & 0.79$\pm$0.01 & 2.29$\pm$0.52 & 17$\pm$1 & 54.5$\pm$2.1 & 54.5$\pm$2.1 & 14.3$\pm$7.0 & 0.15$\pm$0.01 \\
 & 10.5 & 1.23$\pm$0.32 & 37$\pm$14 & 57.8$\pm$0.3 & 57.8$\pm$0.3 & 10.1$\pm$1.9 & 0.74$\pm$0.27 & 2.82$\pm$0.32 & 9$\pm$2 & 49.8$\pm$19.0 & 42.7$\pm$9.0 & 12.4$\pm$5.6 & 0.10$\pm$0.01 \\
\addlinespace[1pt]
Cond.\ inlay & 10.5 & 15.12$\pm$3.29 & 7$\pm$1 & 41.5$\pm$7.5 & 41.5$\pm$7.5 & 10.0$\pm$3.2 & 1.06$\pm$0.10 & 12.79$\pm$0.69 & 4$\pm$1 & 42.9$\pm$11.4 & 38.4$\pm$5.1 & 11.9$\pm$6.7 & 0.42$\pm$0.10 \\
 & 11 & 15.98$\pm$1.35 & 7$\pm$1 & 37.5$\pm$9.3 & 37.5$\pm$9.3 & 7.9$\pm$2.3 & 1.17$\pm$0.20 & 16.87$\pm$0.51 & 4$\pm$0 & 43.5$\pm$7.1 & 43.5$\pm$7.1 & 10.1$\pm$4.9 & 0.61$\pm$0.22 \\
\addlinespace[1pt]
Tuck rib & 10 & 0.54$\pm$0.08 & 119$\pm$43 & 55.1$\pm$2.6 & 51.4$\pm$6.0 & 10.6$\pm$3.2 & 0.80$\pm$0.09 & 1.41$\pm$0.16 & 29$\pm$1 & 48.5$\pm$3.8 & 48.5$\pm$3.8 & 12.5$\pm$0.4 & 0.12$\pm$0.01 \\
 & 10.5 & 0.88$\pm$0.20 & 65$\pm$25 & 55.6$\pm$1.9 & 51.9$\pm$1.4 & 14.5$\pm$1.4 & 0.75$\pm$0.27 & 1.90$\pm$0.07 & 19$\pm$1 & 47.6$\pm$2.6 & 47.6$\pm$2.6 & 11.3$\pm$3.5 & 0.13$\pm$0.01 \\
\midrule
\multicolumn{2}{l}{between/within SD}  & 24.8$\times$ & 7.6$\times$ & 3.0$\times$ & 1.9$\times$ & 1.3$\times$ & 1.5$\times$ & & & & & & \\
\bottomrule
\end{tabular}
\end{table*}
 
\subsection{Tensile Tests}
\label{sec:res_tensile}

Samples were preconditioned with 200 stretch–release cycles. Preconditioning removed most of the initial run‑in of sensitivity: the 50‑cycle rolling mean of GF increased by a median of $24\,\%$ during preconditioning, but changed by only $+3.7$, $+0.9$ and $-0.3\,\%$ within the three subsequent rate blocks.

\textbf{Baseline resistance:} $R_0$ varies across knit structures. Increasing stitch size by one step within each structure (Section~\ref{sec:materials}) raised $R_0$ in five of the seven structures, while other metrics showed no consistent stitch‑size effect. Within each design, $R_0$ continued to increase by $7$ to $14\,\%$
within every rate block. Accordingly, we recalculated $R_0$ on a per-block basis during analysis.

\textbf{Sensitivity:} According to Table \ref{tab:designs}, plain jersey was by far the most sensitive, with GFs of 302 and 329 for stitch size 10.5 and 11, respectively. Among the rib structures, 1×1 rib and the tuck‑enhanced rib reached $65$ to $119$, 2×1 rib $37$ to $52$, and the two cardigan variants (tuck every course) $18$ to $67$. The conductive inlay—where the conductive yarn floats rather than forming loops (i.e., the conductive path is less mechanically engaged)— was the least sensitive (GF around 7) and the only structure with $R_0$ above a few ohms ($15$ to $16\,\Omega$).

\finding{1}{Sensitivity follows the structure family: plain jersey
($\mathrm{GF}\approx 300$) $\gg$ 1x1 rib and tuck rib ($65$ to $119$)
$>$ 2x1 rib and cardigans ($18$ to $67$) $\gg$ conductive inlay
($\approx 7$); the plated loop structures are roughly 3 to 50 times more sensitive
than the floating conductive inlay.}

For each design, we compared GF across the three rate blocks to assess the impact of loading rate on GF. GF was higher at the fast than at the slow loading rate in $31$ of $42$ samples, increasing from an average of $78$ at $3$\,mm/s to $98$ at $14.4$\,mm/s. The impact on other metrics (WR, LR, H) was modest: higher rates shifted them by less than three percentage points.

\textbf{Working Range: }
As shown in Figure~\ref{fig:curves}, all designs exhibit an initial increase in resistance within the first $3\,\%$ strain, followed by a continued rise over most of the $60\,\%$ stroke. The
peak was reached at a median of $57.7\,\%$ strain, and the resistance
fell by at most $17\,\%$ of the peak before the end of the stroke. This behavior is consistent with the contact–separation mechanism described by Jansen et al.~\cite{jansen2020performance}.

\textbf{Linear range:} Across all designs, the response was near-linear within the WR: LR spanned a median of $98\,\%$ of WR and was identical to WR in five designs. Consequently, LR closely mirrored WR’s rankings and offered little additional  discriminative value.

\textbf{Hysteresis:} Hysteresis shows the weakest design dependency ($1.3\times$). $H$ ranged from $4.5\%$ to $14.5\,\%$ strain—lowest for plain jersey (10.5) and highest for tuck rib (10.5). 

\textbf{Cyclic stability:} The detrended cycle-to-cycle standard deviation ranged from $0.68\,\Omega$ (half cardigan, 10.5) to $1.96\,\Omega$ (plain jersey, 10.5), yielding a between-design ratio of $1.5\times$. Detrending removed monotonic drift accounts for $8$--$53\,\%$ of the raw standard-deviation variance (median $27\,\%$), with the largest contribution observed in plain jersey—the most sensitive structure. Thus, cycle-to-cycle scatter is driven more by time-wise drift than by structural differences.

\textbf{Between/within SD: }
To quantify design dependency, the last row of Table~\ref{tab:designs} reports, for each metric, the ratio of the standard deviation across the fourteen design means to the median within-design standard deviation. Ratios well above one indicate that knit structure contributes more variability than sample-to-sample differences. By this measure, $R_0$ and GF vary across designs by $24.8\times$ and $7.6\times$ their within-design standard deviation, respectively. In contrast, WR, LR, H and cyclic stability vary
by only $1.3\times$ to $3.0\times$.

\finding{2}{Sensitivity and baseline resistance are largely determined by knit structure, but WR, LR, H and cyclic stability show only modest between-design differences.}

\subsection{Bending Tests}
\label{sec:res_elbow}

We used two samples per design for the bending test—one previously untested and one that had undergone tensile testing—and report the mean ± standard deviation. After converting joint-angle changes to length changes, the 0–150° stroke corresponds to $57$ to
$66\,\%$ engineering strain (median $61.9\,\%$).

As shown in Figure 4 (dashed lines), all designs exhibited an S‑shaped response: minimal change below~$20\degree$, a steep rise mid‑stroke, and a flattening toward $150\degree$, with the peak at a median of $140\degree$. The initial flat region likely reflects slack take‑up in the mounting before fabric elongation, so the working range began at $3\degree$ to
$13.5\degree$ rather than within the first percent of strain.

\textbf{Same-sample pairing: } For each design, one sample underwent tensile testing followed by bending testing. We compared electromechanical metrics computed at matching strain levels and at the same position within the corresponding rate block. This was repeated for all 14 designs.

\textbf{Design-mean pairing:} For tensile testing, we computed the per-design mean across three samples and compared these means to the metrics of the previously unused sample measured in the bending test

\textbf{Rank-order consistency:} Both pairings yield 14 data points (one per design). Design rankings were compared using the Spearman rank correlation of each metric across the 14 designs for both pairings. For the same-sample pairing, absolute values were also compared via the bending-to-tensile metric ratio (including baseline resistance), summarized as the median ratio across the 14 samples (one sample per design).

\begin{figure}[t]
\centering
\includegraphics[width=\columnwidth]{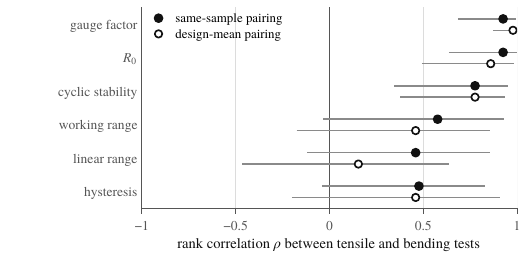}

\caption{Spearman rank correlations between tensile and bending tests, computed per metric for the same‑sample and design‑mean pairings. Whiskers indicate bootstrap 95\,\% confidence intervals across the fourteen designs.} 
\label{fig:cross}
\end{figure}

Figure~\ref{fig:cross} shows the Spearman rank correlation ($\rho$) between rigs for each metric, in both the same-sample and design-mean pairings. GF and $R_0$ achieved high consistency ($\rho = 0.86$ to $0.98$) in both pairings; cyclic stability reached $\rho =0.78$. By contrast, WR, LR, and H exhibited weaker consistency ($\rho = 0.15$ to $0.58$).

\section{Motion sensing performance}
\label{sec:motion}

RQ2 asks whether electromechanical metrics from uniaxial tensile tests reliably predict joint angle estimation error. To evaluate this, we trained machine-learning models to estimate joint angles from each bending-test measurement and then quantified the correlation between the resulting estimation error and the six electromechanical metrics listed in Table~\ref{tab:designs}.

\subsection{Evaluation methodology}
\label{sec:ml_protocol}

We estimate the joint angle as a continuous quantity. For each bending-test session, the resistance time series was segmented into sliding windows of 0.25 s with a 0.125 s step, corresponding to 25 and 12 samples at the 100 Hz rate of our study, comparable to the 50 and 25 samples used at 200 Hz in the reference study~\cite{schepers2026comparative}. Each window was labeled with the mean robot-reported joint angle over the window. Windows within each rate block were split chronologically ($80/20$) into training and testing sets. The three training partitions per session were concatenated to train a single model, while the test partitions were merged to evaluate performance across all three loading/bending rates.

We evaluated joint-angle estimation with the root-mean-squared error (RMSE) and the mean absolute error (MAE) in degrees between the predicted and the measured mean angle of a window, the coefficient of determination $R^2$, and the share of test windows whose prediction lies within $5\degree$ of the measured angle. Predictions were clipped to the $0\degree$--$150\degree$ range of the protocol. RMSE was the primary metric. Uncertainty was quantified via a $95\,\%$ bootstrap confidence interval obtained by resampling the test windows with replacement $10^4$ times, keeping the fitted model fixed.

\textbf{Estimators: } We trained three estimators. All three take the same input: each window is low-pass filtered with a second-order Butterworth filter at $20$\,Hz and standardized with the mean and standard deviation of the training windows. 1) Multi-layer perceptron (MLP): three hidden layers with 128 ReLU units each, optimized with Adam (learning rate $10^{-4}$) and early stopping on a randomly drawn $10\,\%$ validation split from the training windows~\cite{kim2024measurement,hannigan2024distributed}. Because consecutive windows overlap by $50\,\%$, the validation set shares samples with the training set; this affects only the stopping criterion and does not touch the test set. 2) Random forest (RF): 200 trees with a maximum depth of 20~\cite{breiman2001random}. 3) Detrended MLP: the same MLP, with a straight line fitted to the resistance trace of the whole session and subtracted from each filtered window.

We chose standard, window-based estimators rather than temporal sequence models (e.g., LSTMs) to keep the evaluation pipeline relatively transparent and prevent algorithm-level confounds. This choice ensures that differences in estimation error directly reflect the intrinsic electromechanical properties and physical non-idealities of each knit structure.

For each estimator, we quantified the relationship between the six electromechanical metrics (Table~\ref{tab:designs}) and RMSE using Spearman rank correlation across  fourteen designs, reporting the corresponding $p$-value and a bootstrap $95\,\%$ confidence interval over designs. Interpretation focuses on the consistency of correlation signs across classifiers.

\subsection{Estimation Error}
\label{sec:ml_results}

\begin{table*}[t]
\caption{Comparison of joint-angle estimation error across 7 knit structures at two stitch sizes.}
\label{tab:ml}
\centering\footnotesize\setlength{\tabcolsep}{4pt}
\begin{tabular}{ll|rrrrrr|rrr}
\toprule
 & & \multicolumn{6}{c|}{Not previously tensile tested sample ($n=1$)} & \multicolumn{3}{c}{Tensile-tested sample ($n=1$)} \\
 & & \multicolumn{4}{c}{MLP} & Detr.\ MLP & RF & MLP & Detr.\ MLP & RF \\
Structure & Stitch size & RMSE [95\,\% CI] & MAE & $R^2$ & $\leq 5\degree$ & RMSE & RMSE & RMSE & RMSE & RMSE \\
\midrule
Plain Jersey & 10.5 & 10.6 [10.3, 11.0] & 8.2 & 0.941 & 43\% & 10.3 & 8.5 & 11.2 & 11.4 & 9.7 \\
 & 11 & 10.4 [10.1, 10.7] & 7.9 & 0.944 & 38\% & 10.5 & 8.3 & 8.9 & 9.4 & 6.4 \\
\addlinespace[1pt]
1$\times$1 Rib & 10 & 8.5 [8.2, 8.8] & 5.7 & 0.962 & 59\% & 8.7 & 7.8 & 8.4 & 8.2 & 8.1 \\
 & 10.5 & 10.0 [9.6, 10.3] & 7.1 & 0.948 & 53\% & 11.1 & 8.6 & 15.3 & 15.5 & 13.3 \\
\addlinespace[1pt]
Full Cardigan & 10 & 7.7 [7.4, 8.0] & 5.6 & 0.969 & 58\% & 9.1 & 6.8 & 9.8 & 9.9 & 7.4 \\
 & 10.5 & 11.3 [10.9, 11.6] & 8.4 & 0.933 & 44\% & 11.7 & 11.0 & 7.9 & 7.5 & 7.0 \\
\addlinespace[1pt]
Half Cardigan & 10 & 9.2 [8.9, 9.5] & 7.1 & 0.955 & 47\% & 9.8 & 8.4 & 12.6 & 12.7 & 11.9 \\
 & 10.5 & 11.7 [11.2, 12.1] & 7.8 & 0.929 & 52\% & 12.3 & 10.8 & 11.1 & 11.4 & 9.5 \\
\addlinespace[1pt]
2$\times$1 Rib & 10 & 5.5 [5.3, 5.7] & 3.9 & 0.984 & 72\% & 5.8 & 4.9 & 9.1 & 9.2 & 7.2 \\
 & 10.5 & 8.5 [8.2, 8.7] & 6.0 & 0.962 & 61\% & 8.5 & 7.2 & 9.3 & 9.1 & 7.9 \\
\addlinespace[1pt]
Cond.\ inlay & 10.5 & 13.4 [12.9, 13.8] & 9.7 & 0.905 & 39\% & 13.8 & 11.9 & 10.6 & 10.5 & 8.7 \\
 & 11 & 12.5 [12.1, 12.9] & 8.7 & 0.917 & 49\% & 11.5 & 11.9 & 12.3 & 10.8 & 11.8 \\
\addlinespace[1pt]
Tuck Rib & 10 & 9.5 [9.2, 9.8] & 6.9 & 0.952 & 49\% & 9.8 & 6.5 & 7.1 & 7.3 & 5.9 \\
 & 10.5 & 8.3 [8.1, 8.5] & 6.2 & 0.964 & 53\% & 8.5 & 6.1 & 9.5 & 10.0 & 8.4 \\
\midrule
median & & 9.7 & 7.1 & 0.950 & 51\% & 10.1 & 8.3 & 9.6 & 10.0 & 8.2 \\
\bottomrule
\end{tabular}
\end{table*}

We used one preconditioned, previously untested sample per design to compare all 14 designs (seven knit structures × two stitch sizes). Results are summarized in the left-hand columns of Table~\ref{tab:ml}. With the 
MLP, RMSE across the fourteen  samples ranged from $5.5\degree$
(2x1 rib, stitch size 10) to $13.4\degree$ (conductive
inlay, stitch size 10.5), with a median of $9.7\degree$ and a median $R^2$ of $0.95$; around half of the test windows fell within $5\degree$ of the measured angle. The RF achieved a lower RMSE for every design, with a median of $8.3\degree$. Linear detrending changed RMSE by a median of $+0.3\degree$ and improved only two of the fourteen designs.

Error varied far less across designs than sensitivity: a $2.4$-fold spread in RMSE versus a
$49$-fold spread in tensile GF and a $12$-fold spread in bending GF for the same designs (Table~\ref{tab:designs}). The two
conductive inlay designs, the least sensitive on both tensile and bending testing rigs, produced the
two largest errors. The smallest errors were observed for the 2x1 rib
(stitch size 10) and the full cardigan (stitch size 10), both of which rank
mid-pack in sensitivity.

\finding{3}{The estimation error varies only $2.4$-fold across designs
($5.5$ to $13.4\degree$) compared with a 49-fold variation in tensile
sensitivity: the least sensitive sensors are the worst, but the best
performers have intermediate sensitivity.}

We also evaluated one previously tensile-tested sample per design on the bending test rig. As summarized in the right-hand columns of Table~\ref{tab:ml}, these samples yielded a median RMSE of $9.6\degree$, with a median paired difference of $+0.2\degree$ relative to the preconditioned but previously untested samples. Individual paired differences were substantial ($-3.4\degree$ to $+5.4\degree$), and design rankings agreed only weakly between the two sample types ($\rho = 0.27$ for MLP; $\rho = 0.51$ for RF).
These discrepancies may reflect two factors: the prolonged tensile testing (5 hours) prior to bending tests and differences in rate order during the measurement session.

\subsection{Correlation Between Electromechanical Metrics and Estimation Error}
\label{sec:rq2}

\begin{figure}[t]
\centering
\includegraphics[width=\columnwidth]{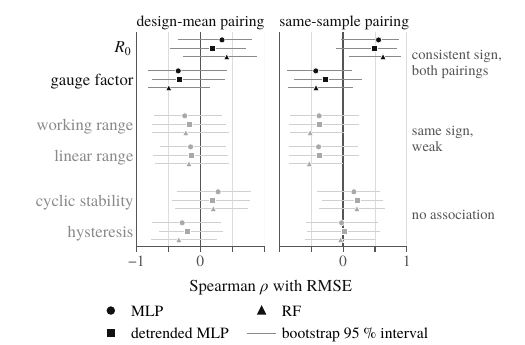}
\caption{Spearman rank correlation $\rho$ between each electromechanical metric and joint-angle estimation error (RMSE), comparing MLP, RF, and detrended MLP under design-mean and same-sample pairings. Whiskers indicate bootstrap 95\,\% intervals across the fourteen designs. Rows are ordered by the median $|\rho|$ over the six estimates (three classifiers x two pairings); the top two
metrics are shown in solid ink. }
\label{fig:rq2}
\end{figure}

Figure~\ref{fig:rq2} shows the Spearman rank correlation between each
electromechanical metric and joint-angle estimation error (RMSE). Two
metrics were associated with the error in every estimator and both
pairings: GF negatively ($\rho = -0.28$ to $-0.49$) and $R_0$ positively
($\rho = +0.19$ to $+0.63$). The association is moderate. The strongest
estimates were $\rho = -0.49$ for GF ($p = 0.07$, design-mean pairing, RF)
and $\rho = +0.63$ for $R_0$ ($p = 0.017$, same-sample pairing, RF), and
only the latter had a bootstrap interval clear of zero. A more sensitive
fabric therefore tends to give a smaller error, but sensitivity explains
little of the spread: the half cardigan at stitch size 10 and the plain
jersey at stitch size 10.5 have GF of $18$ and $302$
(Table~\ref{tab:designs}), yet achieved RMSEs of $8.4\degree$ and
$8.5\degree$ with the RF (Table~\ref{tab:ml}). The positive correlation
with $R_0$ is the same pattern seen from the other side: the two
conductive inlay designs have both the highest $R_0$ and the largest
errors.

WR and LR were also negatively correlated with error in every case, but
weakly ($\rho = -0.15$ to $-0.53$), with intervals that reached zero. H
($|\rho| \leq 0.34$) changed sign between the two pairings, and cyclic
stability reached at most $|\rho| \leq 0.28$. These are the metrics whose
rankings did not transfer between tensile testing and bending testing,
indicating that their tensile values carry little information about joint
performance.

\finding{4}{Of the six electromechanical metrics, only GF and $R_0$ moderately predict joint-angle estimation error; WR and LR do so weakly, and H and CA not at all.}
 
The correlation between sensitivity and joint-angle estimation error is confined to the less-sensitive half of the designs. Among the seven designs with a tensile GF below the median (about $60$) - the two conductive inlays, the two
half cardigans, the full cardigan at stitch size 10.5 and the two 2x1 ribs - 
GF was strongly rank‑correlated with error across all classifiers and both pairings ( $\rho = -0.61$ to $-0.86$).

\finding{5}{Only the less-sensitive designs show a link between tensile sensitivity and joint-angle estimation error, indicating that tensile tests can eliminate poor designs but cannot pinpoint the best.}

For comparability with the reference study~\cite{schepers2026comparative}, we also ran the same three estimators as classifiers over 16 classes of $10\degree$. The two estimators rank the fourteen designs almost identically ($\rho = 0.98$ for the MLP and $1.00$ for the RF); we report the regression results because they directly predict the continuous angle and lower the median error against the measured angle from $11.1\degree$ to $9.7\degree$ with the
MLP and from $9.7\degree$ to $8.3\degree$ with the RF.

\section{Conclusions and Future Work}
We studied how knit structure and stitch size affect electromechanical properties and the extent to which tensile‑test metrics predict in situ joint‑angle error. Across seven structures and two stitch sizes, knit structure chiefly controls sensitivity and baseline resistance; conventional tensile tests provide only partial guidance for sensor design selection. Joint‑relevant evaluation and application‑specific metrics are essential to identify top‑performing designs.

Our study intentionally adopts a focused scope—seven knit structures at two stitch sizes, a single material system and finishing, one joint‑emulation geometry with a linear angle–strain assumption, controlled environmental conditions, and moderate sample sizes—to establish a clean, comparable baseline that clarifies structure‑driven effects. Future work will broaden materials (conductive and companion yarns), refine knit architectures, evaluate durability (wash and aging), and expand kinematics to multiaxial and skin‑influenced motions (shear, twisting, slippage). We also advocate progress toward standardized testing and calibration to enable robust cross‑study comparisons.

\bibliographystyle{IEEEtran}
\bibliography{refs}

\end{document}